\documentclass[sigconf, screen, nonacm]{acmart} 

\usepackage[utf8]{inputenc}
\usepackage{fontenc}
\usepackage{listings}
\usepackage{booktabs}
\usepackage{enumitem}
\usepackage{hyperref}
\usepackage{graphicx}
\usepackage{xcolor}
\usepackage{float}
\usepackage{tikz}
\usetikzlibrary{shapes.geometric, arrows.meta, positioning, shadows, fit, backgrounds, calc, decorations.pathreplacing}

\usepackage{fancyhdr}

\copyrightyear{2026}
\acmYear{2026}
\setcopyright{cc}
\setcctype{by}
\acmConference[WWW Companion '26]{Companion Proceedings of the ACM Web Conference 2026}{April 13--17, 2026}{Dubai, United Arab Emirates}
\acmBooktitle{Companion Proceedings of the ACM Web Conference 2026 (WWW Companion '26), April 13--17, 2026, Dubai, United Arab Emirates}
\acmISBN{979-8-4007-2308-7/2026/04}

\begin{document}

\title[Origin Lens: Privacy-First Image Provenance]{Origin Lens: A Privacy-First Mobile Framework for Cryptographic Image Provenance and AI Detection}
\thanks{This research is funded by the European Union project CyberSecDome under the grant agreement 101120779. This paper was peer-reviewed and accepted at the Future of Information Integrity Research (FIIR) Workshop at The Web Conference 2026 (\url{https://sites.google.com/view/fiirworkshop}); the workshop was subsequently cancelled and the paper was not included in the proceedings. A separate, independent paper on Origin Lens appeared at ACM WebSci '26~\protect\cite{loth2026originlenswebsci}. Origin Lens is part of the first author's doctoral research on AI-generated disinformation~\protect\cite{loth2026symposium}.}

\author{Alexander Loth}
\orcid{0009-0003-9327-6865}
\affiliation{%
  \institution{Frankfurt University of Applied Sciences}
  \city{Frankfurt am Main}
  \country{Germany}
}
\affiliation{%
  \institution{Doctoral Center for Applied Computer Science (PZAI)}
  \city{Wiesbaden}
  \country{Germany}
}
\email{alexander.loth@stud.fra-uas.de}

\author{Dominique Conceicao Rosario}
\email{dominique.conceicaorosario@stud.fra-uas.de}
\author{Peter Ebinger}
\orcid{0009-0002-0108-1802}
\email{peter.ebinger@fra-uas.de}
\author{Martin Kappes}
\orcid{0000-0002-8768-8359}
\email{kappes@fra-uas.de}
\affiliation{%
  \institution{Frankfurt University of Applied Sciences}
  \city{Frankfurt am Main}
  \country{Germany}
}

\author{Marc-Oliver Pahl}
\orcid{0000-0001-5241-3809}
\affiliation{%
  \institution{IMT Atlantique, UMR IRISA, Chaire Cyber CNI}
  \city{Rennes}
  \country{France}
}
\email{marc-oliver.pahl@imt-atlantique.fr}

\renewcommand{\shortauthors}{Alexander Loth et al.}

\begin{abstract}
The proliferation of generative AI poses challenges for information integrity assurance, requiring systems that connect model governance with end-user verification. We present \textbf{Origin Lens}, a privacy-first mobile framework that targets visual disinformation through a layered verification architecture. Unlike server-side detection systems, Origin Lens performs cryptographic image provenance verification and metadata-based AI detection locally on the device via a Rust/Flutter hybrid architecture. Our system integrates multiple signals---including cryptographic provenance, generative-model metadata and watermarks, and optional retrieval-augmented verification---to provide users with graded confidence indicators at the point of consumption. We discuss the framework's alignment with regulatory requirements (EU AI Act, DSA) and its role in verification infrastructure that complements platform-level mechanisms.
\end{abstract}



\maketitle

\pagestyle{fancy}
\fancyhf{}
\fancyhead[L]{\textit{A.\ Loth, D.\ Conceicao Rosario, P.\ Ebinger, M.\ Kappes, M.-O.\ Pahl}}
\fancyhead[R]{\textit{Accepted at the FIIR Workshop, TheWebConf '26}}
\fancyfoot[C]{\thepage}
\renewcommand{\headrulewidth}{0pt}

\section{Introduction}
Generative AI has introduced a \textit{verification asymmetry} in digital societies: while AI models can generate synthetic media at scale, individual users lack tools to assess content authenticity at the point of consumption. Our prior survey on generative AI and fake news documents this dual-use nature of LLMs---enabling new disinformation capabilities while offering potential detection solutions \cite{loth2024blessing}. As noted by Régis et al., the intersection of AI and democratic processes requires technical and governance interventions \cite{regis2025ballot}. This challenge spans multiple dimensions of AI safety research---from \textit{Model Design} (transparency mechanisms like watermarking) to \textit{Model Ecosystem} governance (regulatory compliance and infrastructure).

Existing approaches to misinformation mitigation predominantly rely on centralized, platform-level interventions \cite{gorwa2020platform}. However, this creates dependencies on opaque moderation systems and raises concerns about surveillance and data harvesting \cite{BlochWehba_Surveillance_2021}.

This paper introduces \textbf{Origin Lens}, an open-source, privacy-first mobile framework for cryptographic image provenance and AI detection that builds on the C2PA (Coalition for Content Provenance and Authenticity) standard \cite{c2pa2024spec} alongside heuristic AI detection.\footnote{Source code: \url{https://github.com/aloth/origin-lens}; App Store: \url{https://apps.apple.com/app/id6756628121}; other platforms on the roadmap.} The framework aims to make provenance verification accessible to non-expert users. Our contribution is threefold: (1) a privacy-preserving architecture that performs cryptographic and metadata verification entirely on-device, (2) a defense-in-depth verification pipeline combining cryptographic, heuristic, and contextual signals with graded confidence indicators, and (3) a discussion of how client-side verification tools can complement platform governance.

\section{Related Work}
Recent work in information integrity has focused on robustness evaluation and scalable detection. The \textit{OpenFake} dataset \cite{livernoche2025openfake} provides deepfake benchmarks, while \textit{Veracity} \cite{curtis2025veracity} demonstrates retrieval-augmented fact-checking. Thibault et al.\ audit misinformation detection datasets and evaluation practices \cite{thibault2025guide}.
Technical detection must be paired with effective human-AI interaction. Wang et al.\ show that psychological inoculation has limited effectiveness in a realistic social media feed \cite{lin2025inoculation}. Our JudgeGPT/RogueGPT studies\footnote{\url{https://github.com/aloth/JudgeGPT}; \url{https://github.com/aloth/RogueGPT}} reveal a perception-accuracy gap \cite{loth2026eroding,loth2026collateraleffects}, and empirical analysis shows cognitive fatigue degrades fake content detection by 10.2 percentage points \cite{loth2026collateraleffects}. Puelma Touzel et al.\ \cite{puelmatouzel2025simulating} and Ghafouri et al.\ \cite{ghafouri2024epistemic} observe that information flow complexity and uncertainty communication affect trust calibration.
Article~50(2) of the EU AI Act \cite{euaiact2024} mandates machine-readable marking of AI-generated content, while the Digital Services Act (DSA) \cite{EU_DSA_2022} requires platforms to implement content moderation transparency. Human judges rate text from current LLMs as nearly human-written (mean HumanMachineScores of 0.46--0.50) \cite{loth2026killchaingi}. The C2PA standard \cite{c2pa2024spec} enables decentralized provenance verification, but most tools remain server-dependent. Origin Lens provides a privacy-first, client-side alternative.

\section{System Architecture}
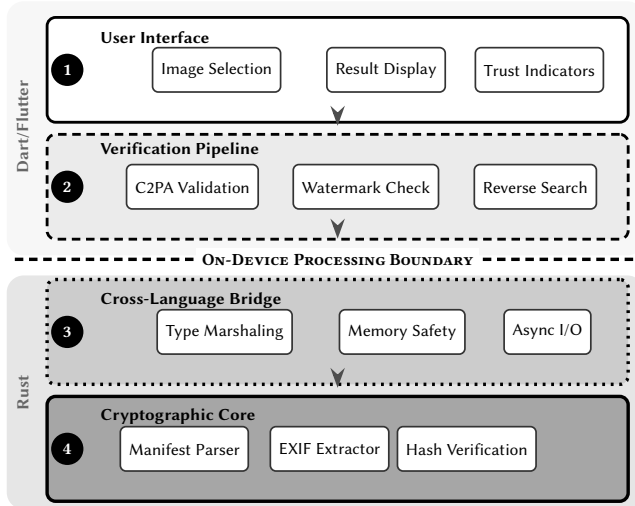
\begin{figure}[htbp]
\centering
\resizebox{\columnwidth}{!}{%
\begin{tikzpicture}[
    layer/.style={
        rectangle,
        rounded corners=3pt,
        minimum width=7.2cm,
        minimum height=1.3cm,
        text centered,
        font=\footnotesize,
        line width=1pt
    },
    ui/.style={layer, fill=white, draw=black},
    service/.style={layer, fill=black!8, draw=black, densely dashed},
    bridge/.style={layer, fill=black!20, draw=black, dotted, line width=1.2pt},
    native/.style={layer, fill=black!35, draw=black, line width=1.2pt},
    arrow/.style={-{Stealth[length=7pt, width=5pt]}, line width=1.5pt, color=black!70, 
                  shorten >=3pt, shorten <=3pt},
    component/.style={
        rectangle,
        rounded corners=2pt,
        minimum height=0.55cm,
        font=\scriptsize\sffamily,
        fill=white,
        draw=black!80,
        line width=0.6pt,
        inner sep=3pt
    },
    layerbadge/.style={
        circle,
        minimum size=0.4cm,
        font=\scriptsize\bfseries,
        fill=black,
        text=white,
        inner sep=1pt
    },
    layerlabel/.style={font=\scriptsize\bfseries\sffamily, text=black}
]

\begin{scope}[on background layer]
    \fill[black!3, rounded corners=6pt] (-4.1, 0.85) rectangle (3.8, -2.3);
    \fill[black!10, rounded corners=6pt] (-4.1, -2.55) rectangle (3.8, -5.45);
\end{scope}

\node[ui] (ui) at (0, 0) {};
\node[layerbadge] at (-3.35, 0) {1};
\node[layerlabel, anchor=west] at (-3.05, 0.4) {User Interface};
\node[component] at (-1.5, 0) {Image Selection};
\node[component] at (0.6, 0) {Result Display};
\node[component] at (2.5, 0) {Trust Indicators};

\node[service] (service) at (0, -1.45) {};
\node[layerbadge] at (-3.35, -1.45) {2};
\node[layerlabel, anchor=west] at (-3.05, -1.0) {Verification Pipeline};
\node[component] at (-1.8, -1.45) {C2PA Validation};
\node[component] at (0.35, -1.45) {Watermark Check};
\node[component] at (2.45, -1.45) {Reverse Search};

\node[bridge] (bridge) at (0, -3.25) {};
\node[layerbadge] at (-3.35, -3.25) {3};
\node[layerlabel, anchor=west] at (-3.05, -2.85) {Cross-Language Bridge};
\node[component] at (-1.4, -3.25) {Type Marshaling};
\node[component] at (0.8, -3.25) {Memory Safety};
\node[component] at (2.6, -3.25) {Async I/O};

\node[native] (native) at (0, -4.7) {};
\node[layerbadge] at (-3.35, -4.7) {4};
\node[layerlabel, anchor=west] at (-3.05, -4.3) {Cryptographic Core};
\node[component] at (-1.9, -4.7) {Manifest Parser};
\node[component] at (-0.1, -4.7) {EXIF Extractor};
\node[component] at (1.6, -4.7) {Hash Verification};

\draw[arrow] (0, -0.65) -- (0, -0.8);
\draw[arrow] (0, -2.1) -- (0, -2.2);
\draw[arrow] (0, -2.5) -- (0, -2.6);
\draw[arrow] (0, -3.9) -- (0, -4.05);

\draw[line width=1.2pt, color=black, densely dashed] 
    (-4.0, -2.35) -- (3.7, -2.35);
\node[font=\scriptsize\scshape\bfseries, fill=white, text=black, inner sep=2pt] at (0, -2.35) {On-Device Processing Boundary};

\node[rotate=90, font=\scriptsize\bfseries\sffamily, text=black!60] at (-3.9, -0.7) {Dart/Flutter};
\node[rotate=90, font=\scriptsize\bfseries\sffamily, text=black!60] at (-3.9, -4.0) {Rust};

\end{tikzpicture}%
}
\caption{Origin Lens architecture: Flutter UI, verification pipeline, FFI bridge, and Rust core. Cryptographic operations execute on-device below the processing boundary.}
\Description{Layered system architecture diagram with four stacked layers (Flutter UI, verification pipeline, cross-language FFI bridge, Rust cryptographic core) connected by downward arrows and separated by an on-device processing boundary.}
\label{fig:architecture}
\end{figure}

Origin Lens employs a hybrid mobile architecture optimized for performance and memory safety (see Figure~\ref{fig:architecture}). The core logic is implemented in \textbf{Rust} \cite{matsakis2014rust} for its memory safety guarantees \cite{Jung_RustBelt_2017} when parsing complex binary formats, while the user interface is built with \textbf{Flutter} \cite{flutter2024} for cross-platform mobile deployment.

\textbf{Defense-in-Depth Strategy.} Given epistemic uncertainty in modern media \cite{ghafouri2024epistemic}, we implement a four-layer pipeline: (1)~\textit{Cryptographic Provenance}---parse JUMBF (JPEG Universal Metadata Box Format) boxes to validate C2PA manifests, verify X.509 trust chains, and compute SHA-256 hashes ensuring hard binding \cite{c2pa2024spec,xie2022proves,bowen2022zkimg}; (2)~\textit{Heuristic Metadata}---analyze EXIF and IPTC photo metadata for generative model artifacts (e.g., Stable Diffusion parameters); (3)~\textit{Watermark Detection}---detect imperceptible watermarks (e.g., SynthID) via API \cite{jiang2023evading,cox1997secure,kwon2025synthidimage}; (4)~\textit{Contextual Verification}---opt-in reverse image search for prior attributions.

\textbf{Privacy and Uncertainty Communication.}
In contrast to cloud-based detection services, Origin Lens processes C2PA and metadata entirely on-device via FFI (Foreign Function Interface) bindings, following Privacy by Design principles \cite{cavoukian2011privacy} with privacy as the default. Cryptographic provenance and heuristic metadata analysis require no network access. Watermark detection (e.g., SynthID) sends the image to an external API and runs automatically when no C2PA manifest is found. Contextual verification (reverse image search) transmits image data to external services, leaving digital traces; this feature is therefore strictly opt-in and disabled by default, aligning with GDPR data protection by design and by default (Article~25) and data minimization (Article~5(1)(c)) \cite{gdpr2016}.

Communicating uncertainty is a known challenge \cite{ghafouri2024epistemic}. Origin Lens implements a hierarchical confidence model: \textit{high} (valid C2PA with trusted root), \textit{medium} (EXIF patterns, watermarks), and \textit{low} (opt-in reverse image search requiring user interpretation).

\section{Security \& Threat Modeling}
We applied STRIDE threat modeling \cite{shostack2014stride} to the Origin Lens verification pipeline, analyzing each architectural layer (Figure~\ref{fig:architecture}) for potential attack vectors.

\textbf{Identified Threats.} At the \textit{Verification Pipeline} layer, we identify two primary risks: (1)~\textit{Manifest Stripping} (Tampering)---adversaries remove C2PA metadata during redistribution, a known limitation of content credentials \cite{c2pa2024spec}; and (2)~\textit{Certificate Spoofing} (Spoofing)---attackers forge X.509 certificates to inject false provenance claims. At the \textit{Cross-Language Bridge}, malformed input could trigger memory corruption \cite{McCormack_ICSE25}; Rust's ownership model mitigates this at the \textit{Cryptographic Core}. The \textit{User Interface} faces Information Disclosure risks if verification results are cached insecurely.

\textbf{Mitigations.} Origin Lens enforces hard binding validation \cite{c2pa2024spec}: SHA-256 hashes bind the manifest to image content---any modification invalidates the credential. Against spoofing, signer certificate chains are validated through the c2pa-rs reference implementation.

\textbf{Open Challenges.} The analog hole---screen capture---circumvents cryptographic provenance \cite{Radharapu_RealSeal_2024}. We address this through defense-in-depth: heuristic metadata and watermark detection provide secondary signals when manifests are absent. Ecosystem adoption remains critical for verification coverage \cite{c2pa2024spec}.

\section{Evaluation \& Discussion}
On iOS (iPhone 15 Pro), C2PA validation completes in under 500\,ms for 12\,MP images, with EXIF parsing under 50\,ms.
This latency is acceptable for interactive use.
For result communication, Origin Lens uses a traffic-light UI \cite{wickens1999traffic, Stojkovski_TrafficLight_2021}: \textit{green} (valid C2PA from trusted root), \textit{purple} (C2PA/EXIF indicating generative origin), \textit{red} (hash mismatch or broken chain), and \textit{orange} (no manifest found, expired certificate, or parsing issue).

\textbf{Limitations.} C2PA effectiveness depends on ecosystem adoption \cite{c2pa2024spec}. Adversarial actors may employ manifest stripping or analog-hole attacks. Heuristic detection faces distribution shift as models evolve \cite{thibault2025guide,verdoliva2020mediaforensics,guo2025deepfakeforensics,cozzolino2020noiseprint,mareen2023comprint,lukas2006digital}, and demographic predictors show weaker effects for AI-generated content \cite{loth2026collateraleffects}. Ultimately, technical tools require a complementary culture of verification and data literacy to achieve societal impact \cite{loth2021decisively,qian2023fighting}.

\section{Conclusion and Future Directions}
Origin Lens provides an open-source, privacy-first implementation for on-device image provenance verification. By performing verification locally with graded uncertainty signals, the framework complements platform-level mechanisms.
Future work includes lightweight neural networks for pixel-based detection, privacy-preserving federated aggregation, and cross-jurisdictional provenance standards.
As our research continues, we invite experts to participate in our survey on verification practices \cite{loth2026verification}: \url{https://github.com/aloth/verification-crisis}.

\balance
\bibliographystyle{ACM-Reference-Format}
\bibliography{references}


\clearpage

\appendix

\setcounter{page}{1}
\renewcommand{\thepage}{S\arabic{page}}
\renewcommand{\thefigure}{S\arabic{figure}}
\renewcommand{\thetable}{S\arabic{table}}
\setcounter{figure}{0}
\setcounter{table}{0}

\begin{center}
    \LARGE \textbf{Supplementary Materials}
\end{center}

This supplement provides additional figures, architectural details, user interface screenshots, and regulatory context that support the main text.

\section{Defense-in-Depth Verification Pipeline}
\label{sec:supp-pipeline}

Origin Lens implements a four-layer defense-in-depth strategy for image verification, where each layer provides independent verification signals with decreasing confidence levels:

\textbf{Layer 1: C2PA Provenance.} The primary verification layer parses JUMBF-embedded C2PA manifests, validates X.509 certificate chains, and verifies SHA-256 hard bindings between the manifest and image content.

\textbf{Layer 2: EXIF/IPTC Metadata.} When C2PA manifests are absent, the system analyzes EXIF and IPTC metadata for generative AI signatures, including Stable Diffusion parameters, DALL-E identifiers, and Midjourney tags.

\textbf{Layer 3: Watermark Detection.} Detection of imperceptible watermarks (e.g., Google SynthID) via an external API, run automatically when no C2PA manifest is found, provides additional signals for AI-generated content identification.

\textbf{Layer 4: Contextual Verification.} As a final fallback, users may opt in to reverse image search for prior attributions and contextual information.

\begin{figure}[!b]
\centering
\resizebox{\columnwidth}{!}{%
\begin{tikzpicture}[
  layer/.style={draw=gray!50, rounded corners=5pt, 
                minimum height=1.0cm, align=center, font=\small},
]
  \node[layer, fill=black!25, minimum width=9cm] (c2pa) at (0,2.2) 
    {\textbf{1. C2PA Provenance} (Primary)\\Cryptographic Signature Verification};
  
  \node[layer, fill=black!18, minimum width=7.5cm] (exif) at (0,0.9) 
    {\textbf{2. EXIF Metadata} (Secondary)\\AI Generator Signatures};
  
  \node[layer, fill=black!12, minimum width=6cm] (synth) at (0,-0.3) 
    {\textbf{3. SynthID} (Tertiary)\\Invisible Watermarks};
  
  \node[layer, fill=black!6, minimum width=4.5cm] (context) at (0,-1.5) 
    {\textbf{4. Reverse Search} (Quaternary)\\Context Verification};
\end{tikzpicture}%
}
\caption{Defense-in-depth verification layers. Primary layers (top) provide highest confidence; lower layers offer fallback signals when cryptographic provenance is unavailable.}
\Description{Inverted-pyramid diagram with four stacked layers of decreasing width: C2PA Provenance (top, widest), EXIF Metadata, SynthID watermarks, and Reverse Search (bottom, narrowest).}
\label{fig:supp-defense}
\end{figure}
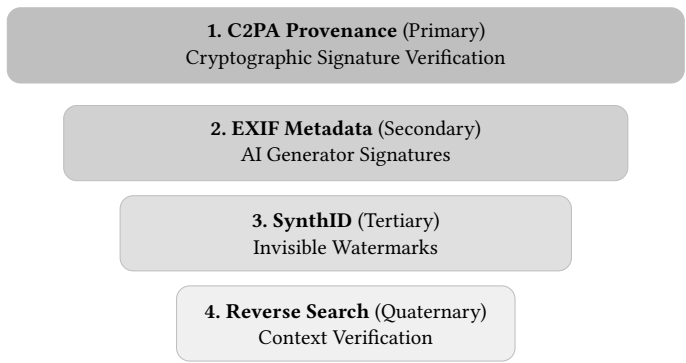

\noindent Figure~\ref{fig:supp-defense} illustrates this layered approach.

\section{C2PA Manifest Structure}
\label{sec:supp-c2pa}

The C2PA standard defines a hierarchical manifest structure embedded within image files, consisting of four core components:

\textbf{Assertions} contain metadata about the content's creation, including timestamps, software used, and editing actions performed.

\textbf{Claims} aggregate assertions and establish relationships with ingredient manifests (for composite images).

\textbf{Signatures} use ECDSA or RSA algorithms with X.509 certificates to cryptographically bind claims to the signing entity.

\textbf{Hard Bindings} compute SHA-256 hashes over the asset's bytes (excluding the embedded manifest), ensuring any modification invalidates the manifest.

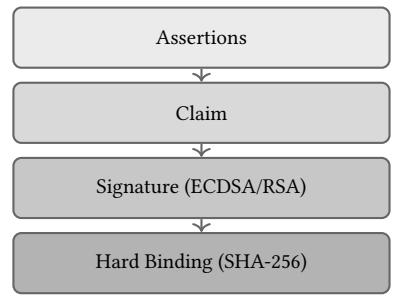
\begin{figure}[!b]
\centering
\begin{tikzpicture}[
  box/.style={draw=black!60, fill=white, rounded corners=3pt,
              minimum width=5cm, minimum height=0.8cm,
              font=\small, line width=0.8pt}
]
  \node[box, fill=black!8] (assert) at (0,2.4) {Assertions};
  \node[box, fill=black!15] (claim) at (0,1.4) {Claim};
  \node[box, fill=black!22] (sig) at (0,0.4) {Signature (ECDSA/RSA)};
  \node[box, fill=black!30] (hash) at (0,-0.6) {Hard Binding (SHA-256)};
  
  \draw[->, thick, black!60] (assert) -- (claim);
  \draw[->, thick, black!60] (claim) -- (sig);
  \draw[->, thick, black!60] (sig) -- (hash);
\end{tikzpicture}
\caption{C2PA manifest structure showing the relationship between assertions, claims, cryptographic signatures, and content hash bindings.}
\Description{Diagram showing the C2PA manifest structure with four stacked layers: Assertions, Claim, Signature (ECDSA/RSA), and Hard Binding (SHA-256), connected by arrows.}
\label{fig:supp-c2pa}
\end{figure}

\noindent Figure~\ref{fig:supp-c2pa} illustrates these relationships.

\section{System Architecture Details}
\label{sec:supp-architecture}

Origin Lens employs a hybrid mobile architecture optimized for performance and memory safety, separating the Dart/Flutter presentation layer from the Rust cryptographic core. The architecture provides several benefits:

\begin{itemize}
  \item \textbf{Memory Safety:} Rust's ownership model prevents buffer overflows and use-after-free vulnerabilities when parsing complex binary formats.
  \item \textbf{Performance:} Native code execution for computationally intensive cryptographic operations.
  \item \textbf{Cross-Platform:} Flutter enables deployment to iOS, Android, and desktop platforms from a single codebase.
\end{itemize}

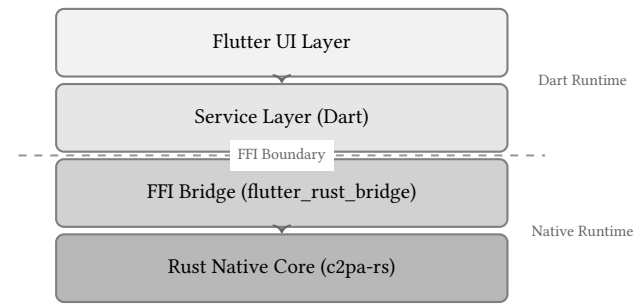
\begin{figure}[!b]
\centering
\begin{tikzpicture}[
  box/.style={draw=black!50, fill=white, rounded corners=3pt,
              minimum width=6cm, minimum height=0.9cm,
              font=\small, line width=0.8pt}
]
  \node[box, fill=black!5] (ui) at (0,2.4) {Flutter UI Layer};
  \node[box, fill=black!10] (svc) at (0,1.4) {Service Layer (Dart)};
  \node[box, fill=black!18] (ffi) at (0,0.4) {FFI Bridge (flutter\_rust\_bridge)};
  \node[box, fill=black!28] (rust) at (0,-0.6) {Rust Native Core (c2pa-rs)};
  
  \draw[->, thick, black!60] (ui) -- (svc);
  \draw[->, thick, black!60] (svc) -- (ffi);
  \draw[->, thick, black!60] (ffi) -- (rust);
  
  \node[font=\scriptsize, text=black!60] at (4,1.9) {Dart Runtime};
  \node[font=\scriptsize, text=black!60] at (4,-0.1) {Native Runtime};
  
  \draw[dashed, black!40, line width=0.8pt] (-3.5,0.9) -- (3.5,0.9);
  \node[font=\scriptsize, fill=white, text=black!60] at (0,0.9) {FFI Boundary};
\end{tikzpicture}
\caption{Origin Lens hybrid architecture. The FFI boundary separates managed Dart code from native Rust, enabling memory-safe cryptographic operations.}
\Description{Layered architecture diagram showing four stacked components: Flutter UI Layer, Service Layer (Dart), FFI Bridge, and Rust Native Core (c2pa-rs), with arrows indicating data flow.}
\label{fig:supp-arch}
\end{figure}

\noindent Figure~\ref{fig:supp-arch} presents the detailed layered architecture.

\section{Analysis Workflow}
\label{sec:supp-workflow}

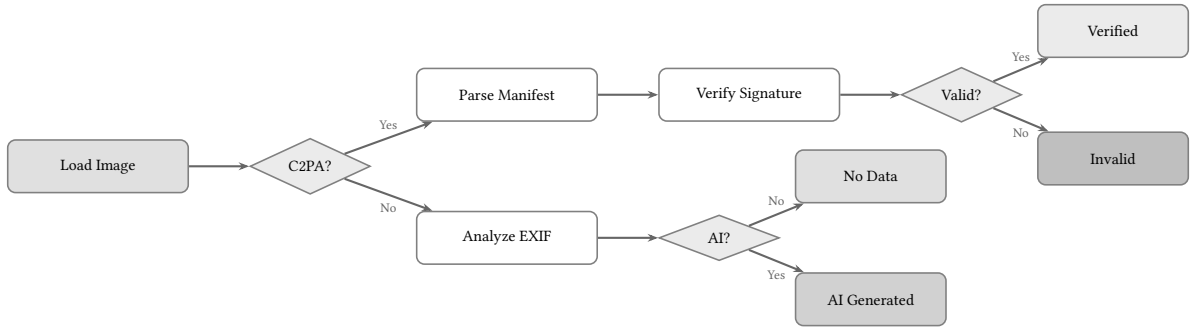
\begin{figure*}[t]
\centering
\begin{tikzpicture}[
  node distance=0.8cm,
  box/.style={draw=black!50, fill=white, rounded corners=3pt,
              minimum width=2.4cm, minimum height=0.7cm, font=\scriptsize,
              line width=0.6pt},
  decision/.style={diamond, aspect=2, draw=black!50, fill=black!8,
                   minimum width=1.6cm, font=\scriptsize, line width=0.6pt},
  result/.style={box, minimum width=2cm},
  arrow/.style={-{Stealth[length=4pt]}, thick, black!60}
]
  \node[box, fill=black!12] (start) {Load Image};
  \node[decision, right=of start] (c2pa) {C2PA?};
  
  \node[box, above right=0.4cm and 1cm of c2pa] (parse) {Parse Manifest};
  \node[box, right=of parse] (verify) {Verify Signature};
  \node[decision, right=of verify] (valid) {Valid?};
  \node[result, fill=black!8, above right=0.3cm and 0.6cm of valid] (ok) {Verified};
  \node[result, fill=black!25, below right=0.3cm and 0.6cm of valid] (fail) {Invalid};
  
  \node[box, below right=0.4cm and 1cm of c2pa] (exif) {Analyze EXIF};
  \node[decision, right=of exif] (ai) {AI?};
  \node[result, fill=black!18, below right=0.3cm and 0.6cm of ai] (aiyes) {AI Generated};
  \node[result, fill=black!12, above right=0.3cm and 0.6cm of ai] (nodata) {Warning};
  
  \draw[arrow] (start) -- (c2pa);
  \draw[arrow] (c2pa) -- node[above, font=\tiny] {Yes} (parse);
  \draw[arrow] (c2pa) -- node[below, font=\tiny] {No} (exif);
  \draw[arrow] (parse) -- (verify);
  \draw[arrow] (verify) -- (valid);
  \draw[arrow] (valid) -- node[above, font=\tiny] {Yes} (ok);
  \draw[arrow] (valid) -- node[below, font=\tiny] {No} (fail);
  \draw[arrow] (exif) -- (ai);
  \draw[arrow] (ai) -- node[above, font=\tiny] {No} (nodata);
  \draw[arrow] (ai) -- node[below, font=\tiny] {Yes} (aiyes);
\end{tikzpicture}
\caption{Image analysis workflow showing decision points and four possible verification outcomes.}
\Description{Flowchart showing the image analysis process: Load Image leads to C2PA check, branching to manifest parsing and signature verification (Yes path) or EXIF analysis and AI detection (No path), with four possible outcomes: Verified, Invalid, AI Generated, or Warning.}
\label{fig:supp-flow}
\end{figure*}

The image analysis workflow follows a decision tree that first checks for C2PA manifests and falls back to heuristic analysis when cryptographic provenance is unavailable. Images with C2PA manifests undergo cryptographic verification (signature and hash validation), while images without manifests are analyzed for AI generation signatures in EXIF metadata. The workflow produces four possible outcomes: Verified, Invalid, AI Generated, or Warning. Figure~\ref{fig:supp-flow} illustrates this complete workflow.

\section{Verification Status Indicators}
\label{sec:supp-status}

Origin Lens communicates verification results using a traffic-light-inspired visual system. Table~\ref{tab:supp-status} describes each status indicator and its meaning.

\begin{table}[!htbp]
\centering
\caption{Verification status indicators and their meanings.}
\label{tab:supp-status}
\begin{tabular}{llp{4.4cm}}
\toprule
\textbf{Status} & \textbf{Color} & \textbf{Description} \\
\midrule
Verified & Green & Valid C2PA manifest with trusted certificate chain \\
AI Generated & Purple & Content identified as AI-generated via C2PA or EXIF markers \\
Warning & Orange & No manifest found, expired certificate, or parsing issue \\
Invalid & Red & Hash mismatch, broken certificate chain, or detected manipulation \\
\bottomrule
\end{tabular}
\end{table}

\section{Regulatory Alignment}
\label{sec:supp-regulatory}

Origin Lens aligns with emerging EU regulations on AI transparency and cybersecurity. Table~\ref{tab:supp-regulatory} summarizes how the framework addresses relevant regulatory requirements.

\begin{table*}[!htbp]
\centering
\caption{Regulatory alignment of Origin Lens features.}
\label{tab:supp-regulatory}
\begin{tabular}{lp{6cm}p{6cm}}
\toprule
\textbf{Regulation} & \textbf{Relevant Requirements} & \textbf{Origin Lens Alignment} \\
\midrule
EU AI Act (2024/1689) & Article 50(2): Machine-readable marking of AI-generated content & C2PA manifest parsing and AI generation detection via metadata \\
GDPR (2016/679) & Article 25: Privacy by Design; Article 5(1)(c): Data minimization & On-device processing; no server transmission for core verification \\
Cyber Resilience Act (2024/2847) & Security-by-design for digital products & Rust memory safety; X.509 certificate validation \\
NIS2 Directive (2022/2555) & Supply chain security requirements & Local trust store; certificate chain verification \\
\bottomrule
\end{tabular}
\end{table*}

\section{User Interface Screenshots}
\label{sec:supp-screenshots}

\begin{figure*}[!htbp]
\centering
\begin{tabular}{@{}ccc@{}}
\includegraphics[width=0.24\textwidth]{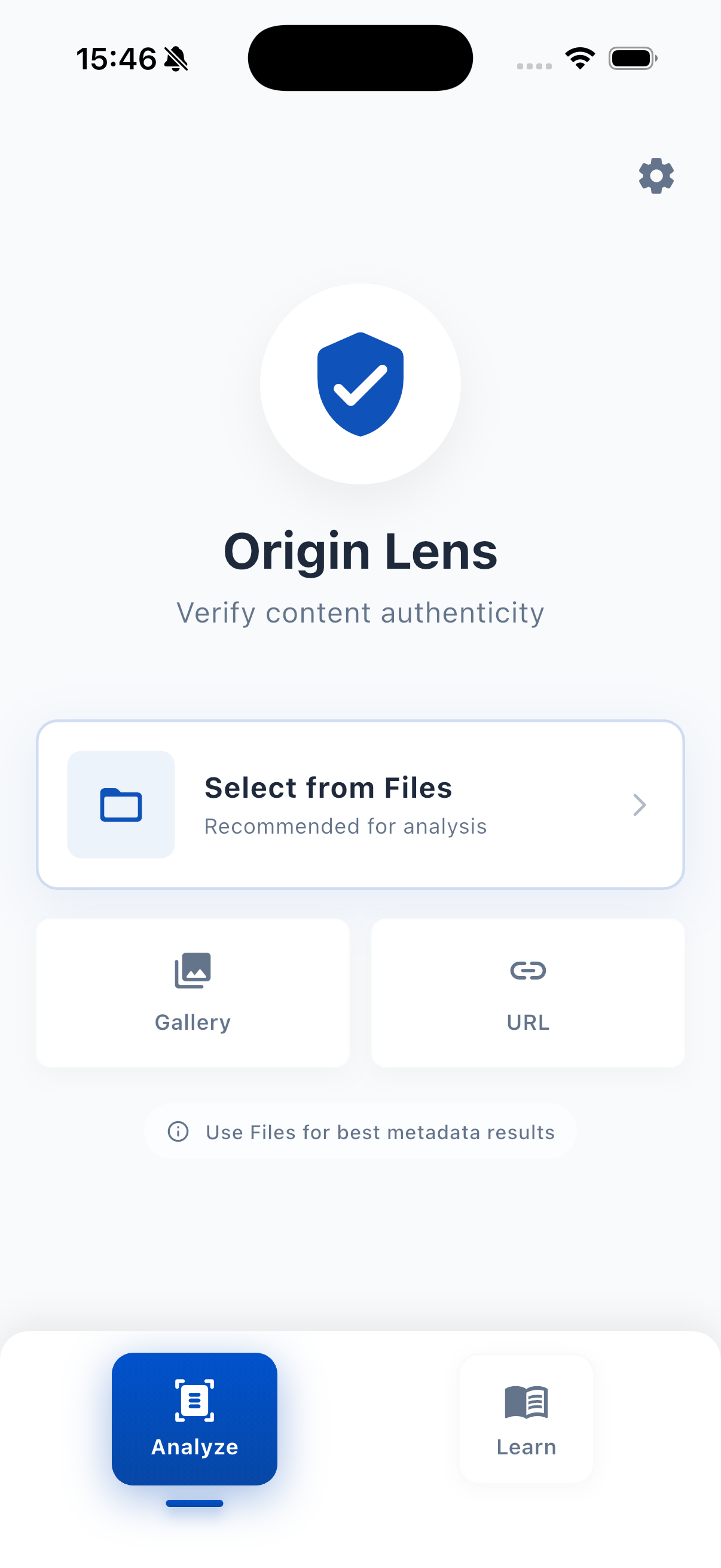} &
\includegraphics[width=0.24\textwidth]{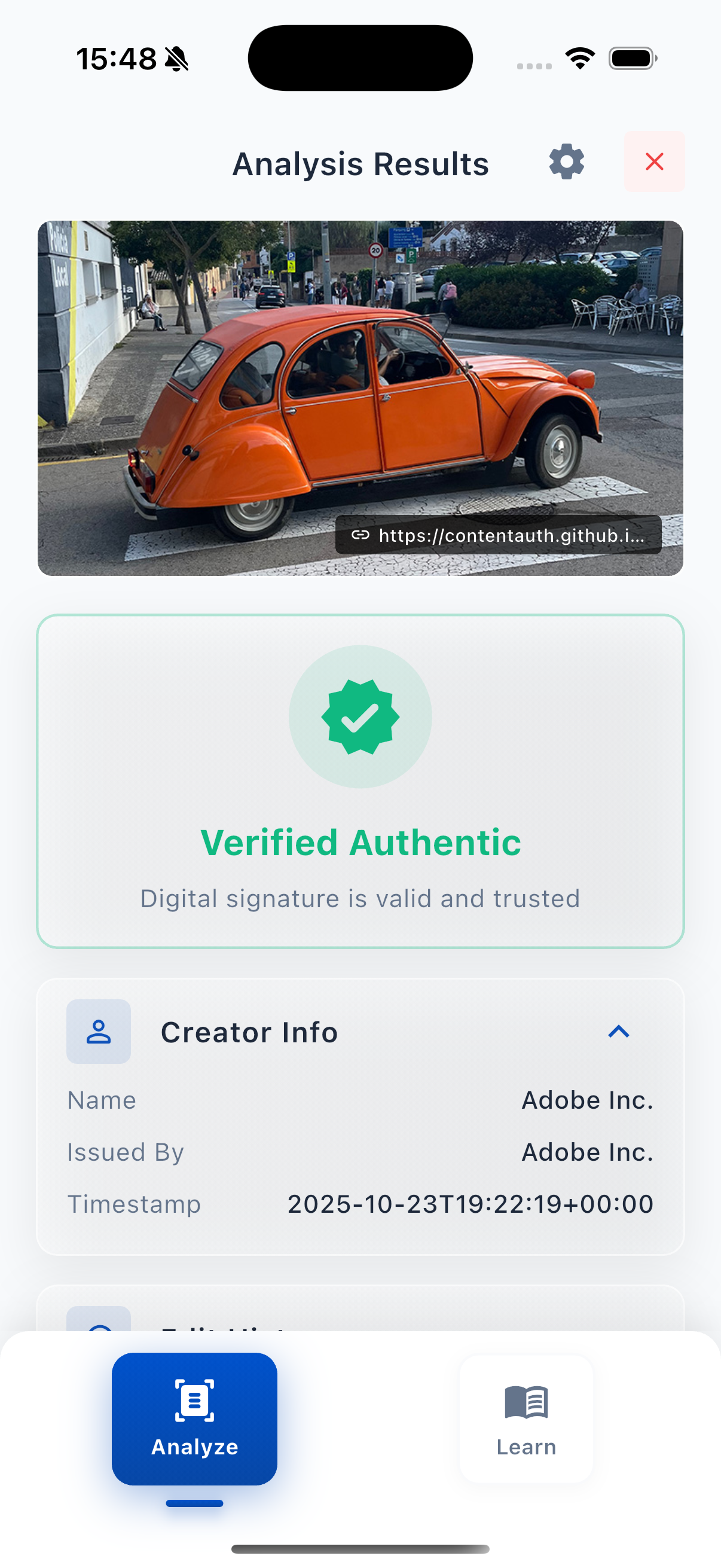} &
\includegraphics[width=0.24\textwidth]{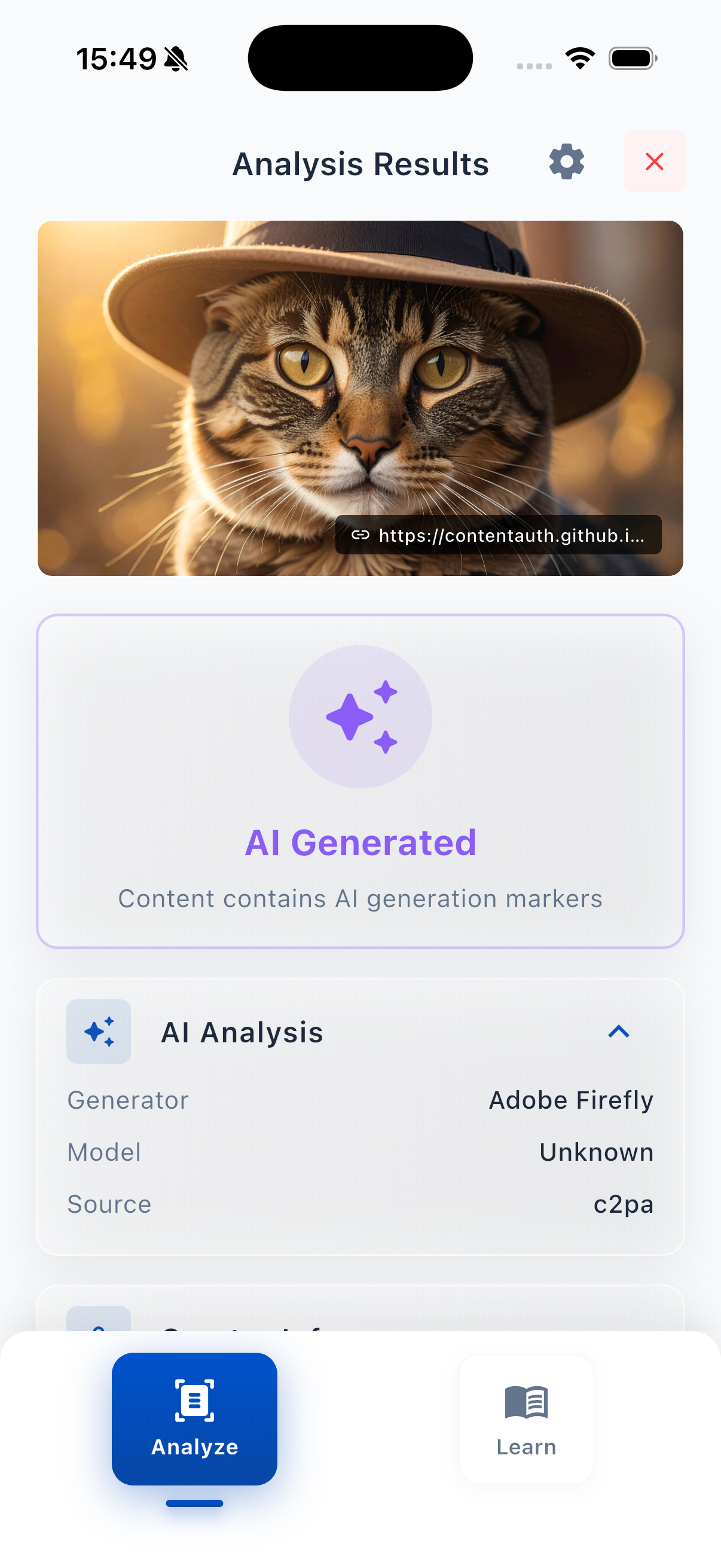} \\
(a) Dashboard & (b) Verified & (c) AI Generated \\[1em]
\includegraphics[width=0.24\textwidth]{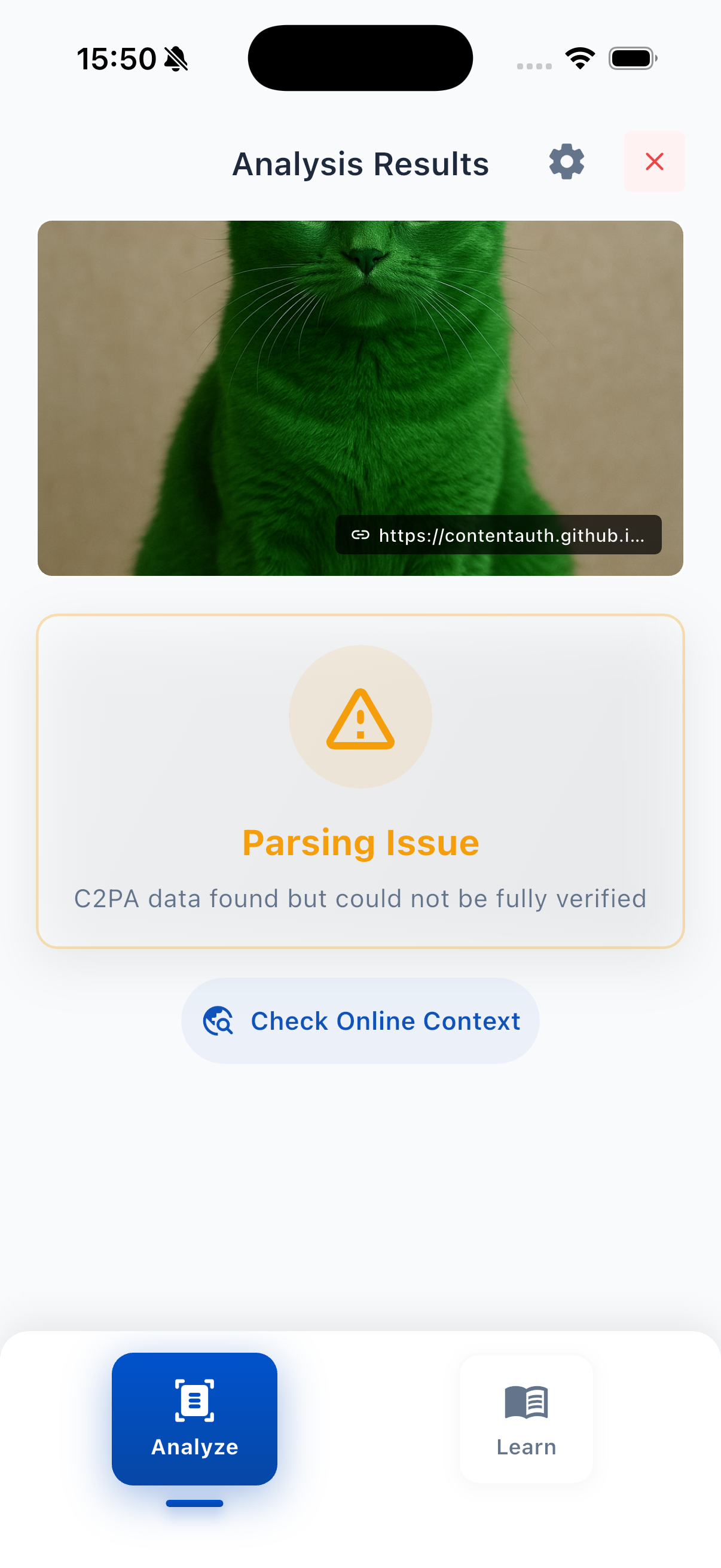} &
\includegraphics[width=0.24\textwidth]{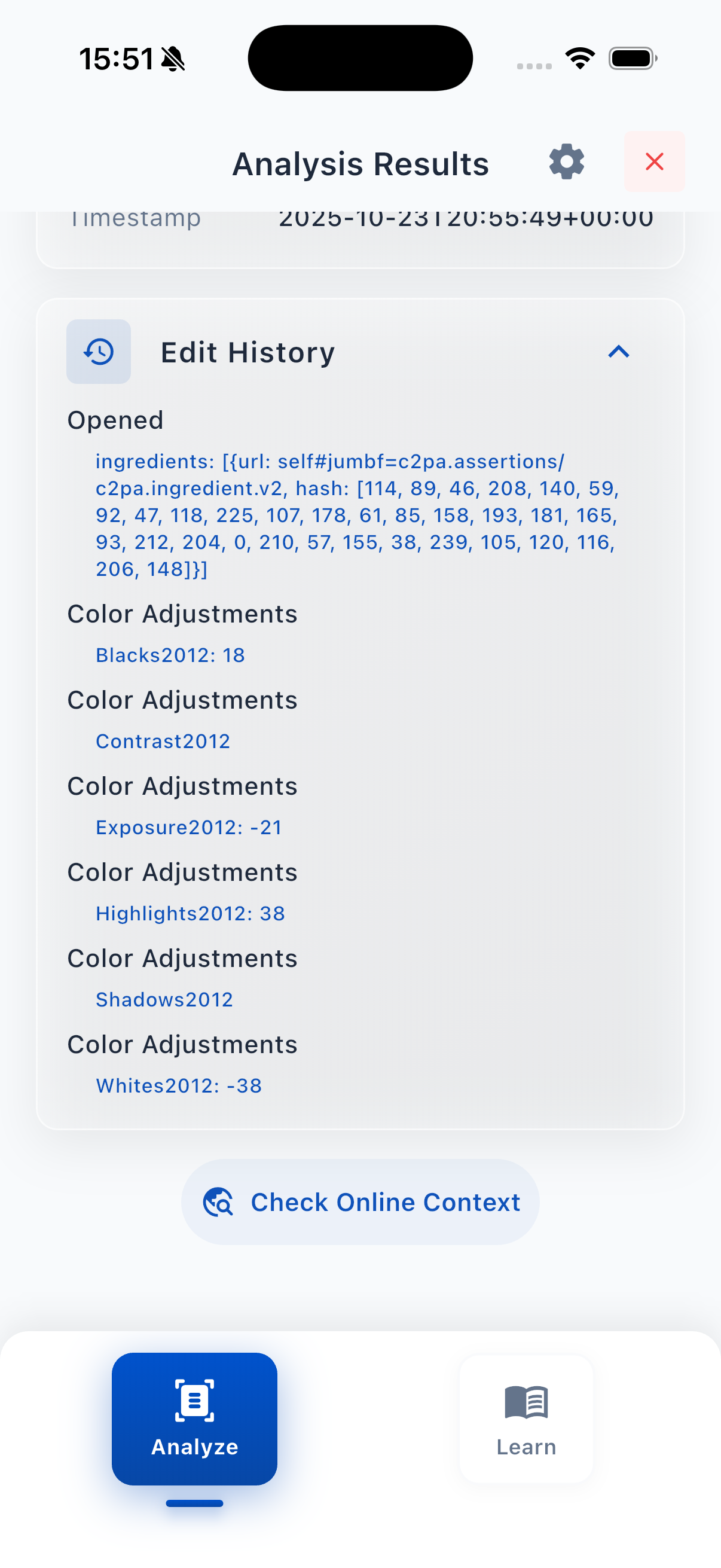} &
\includegraphics[width=0.24\textwidth]{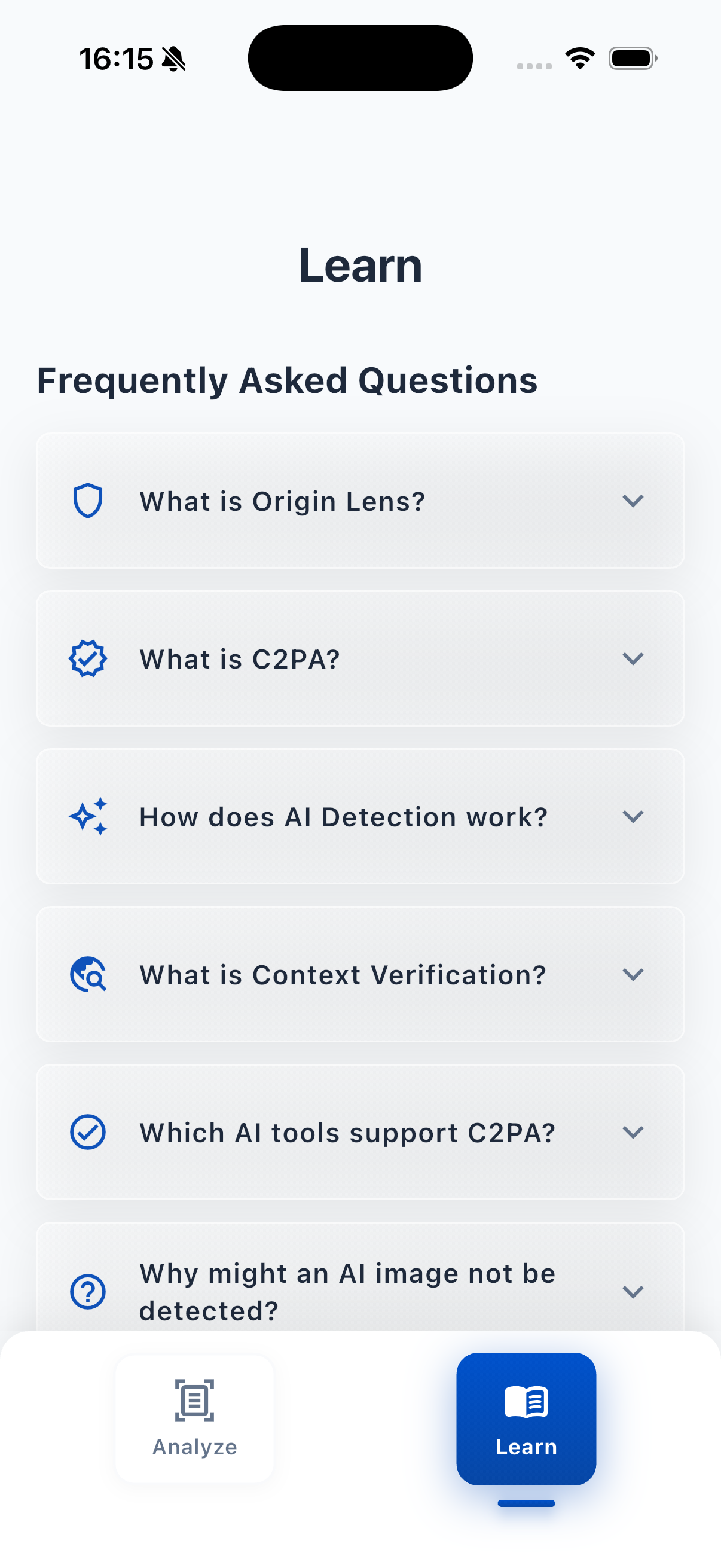} \\
(d) Parsing Issue & (e) Edit History & (f) FAQ / Learn \\
\end{tabular}
\caption{Origin Lens user interface screenshots: (a) Dashboard with upload options (Files, Gallery, URL), (b) Verified result with valid C2PA manifest and trusted signature chain, (c) AI-generated content detected via Adobe Firefly metadata, (d) Parsing issue warning when signature chain is incomplete, (e) Detail view showing C2PA manifest history with timestamps and ingredient hashes, (f) Learn section with expandable FAQ on C2PA and AI detection.}
\Description{Six mobile app screenshots arranged in a 3x2 grid showing the Origin Lens interface: dashboard, verified result, AI-generated detection, parsing issue warning, edit history details, and FAQ section.}
\label{fig:supp-screenshots}
\end{figure*}

This section presents the Origin Lens user interface across different verification scenarios. Figure~\ref{fig:supp-screenshots} shows: (a) the main dashboard with upload options, (b) a verified result with valid C2PA manifest, (c) AI-generated content detection, (d) a parsing issue warning, (e) the detailed manifest history view, and (f) the educational FAQ section.

\end{document}